\documentclass[11pt]{article}
\usepackage{amssymb}
\usepackage{amsmath}
\usepackage[mathscr]{euscript}
\usepackage{graphicx}

\makeatletter  
  
\@addtoreset{equation}{section} 
\makeatother 
\def\Slash#1{/\hspace{-0.23cm}{#1}} 
\newcommand{\del}{\partial}
\def\beq{\begin{eqnarray}}
\def\eeq{\end{eqnarray}}

\def\ln{\mbox{ln}}

\def\befc{\begin{figure}[h]\begin{center}}
\def\eefc{\end{center}\end{figure}}

\def\bra{\langle}    \def\ket{\rangle}

\def\feynint2{\int_0^12xdx\int_0^1dy}

\def\mn{M_N}         
\def\ms{M_S}
\def\ma{M_A}
\def\mq{m_q}

\def\tr{\,\hbox{tr}\,}   
\def\calA{{\cal A}}

    \def\calS{{\cal S}}

\def\calL{{\cal L}}

\begin{document}

\title{{\bf The mass of the nucleon  in a chiral quark-diquark model}}
\author{Keitaro Nagata$^1$, Atsushi Hosaka$^1$ and Laith. J. Abu-Raddad$^2$\\ 
1 {\it Research Center for Nuclear Physics (RCNP), Osaka university,}\\
{\it Ibaraki 567-0047, Japan}\\
2 {\it Department of Infectious Disease Epidemiology, Imperial
College London}\\ {\it  St Mary's Campus, Norfolk Place,
 London W2 1PG, United Kingdom}}
\date{}
\maketitle 
\begin{center}
{\bf Abstract}
\end{center}
{\small
 
The mass of the nucleon is studied in a chiral quark-diquark
model. Both scalar and axial-vector diquarks are taken into
account for the construction of the nucleon state. 
After the hadronization procedure to obtain an effective 
meson-baryon Lagrangian, the quark-diquark self-energy is calculated
in order to generate the baryon kinetic term as well as 
 the mass of the nucleon.
It turns out that 
both the scalar and axial-vector parts of the self-energy are
attractive for the mass of the nucleon.
We investigate the range of parameters that can reproduce 
the mass of the nucleon.}
\section{Introduction}

 An effective Lagrangian approach is an useful method for the
description of hadron properties at low energies.
Such a Lagrangian contains
various terms and parameters expressing not only structures of mesons
and baryons but also their interactions.
A microscopic description for such terms is desired, especially
when we consider, for instance,
character changes of hadrons at finite temperatures and
densities, which is one of the interesting topics of current
hadron physics.

Eventually, QCD should address this issue, but the present situation is not
very satisfactory.
If we start, however, from an intermediate QCD oriented
theory, we can make a reasonably good achievement.
One of such approach is the Nambu-Jona-Lasinio
model~\cite{NJL,Hatsuda:1994pi,Vogl:1991qt} for mesons, 
and the quark diquark model for mesons and 
baryons~\cite{Ebert:1997hr,Abu-Raddad:2002pw}.
The models have been tested to a great extent for the description
of various meson and baryon properties.
It was then shown that the hadronization method based
on the path-integral
formalism is useful, because it can incorporate hadron
structure in terms of quarks and diquarks with respecting
important symmetries such as the gauge and chiral symmetries.
This idea was first investigated by Cahill~\cite{Cahill:1988zi} and
Reinhardt~\cite{Reinhardt:1989rw}.
This, then, was also investigated by Ebert and Jurke in a
simplified framework~\cite{Ebert:1997hr}, which was later more elaborated by
Abu-Raddad et al~\cite{Abu-Raddad:2002pw}.
Recently, the method was applied also to the nuclear force by
the present authors~\cite{Nagata:2003gg}.
Nonetheless, these previous studies were done only with the scalar diquark,
though the construction of the baryon
requires two types of diquarks: scalar and axial-vector ones.
The inclusion of the axial-vector diquarks is crucially important for 
the description of spin-isospin quantities such as the
axial coupling constant $g_A$ and isovector magnetic moment $\mu$ of the
nucleon, and also the nuclear force.

In this paper, we extend our previous study and calculate the
nucleon mass with the inclusion of the axial-vector diquark.
This is a necessary step to complete the program of the
hadronization method. It is shown that by choosing suitable
parameters, the mass of the nucleon is reproduced with the same
significant amount of the axial-vector diquark component, which will help
improve the observables such as $g_A$ and isovector magnetic moment.   

The paper is organized as follows. In section 2, we construct a
microscopic (quark-diquark) Lagrangian and derive the
macroscopic (meson-baryon) Lagrangian through the hadronization of the
microscopic Lagrangian. In section 3, we study the quark-diquark
self-energy and calculate the mass of the nucleon. In section 4,
 we present numerical results. 
The final section is devoted to summary and conclusions.
\section{Lagrangian}

We briefly review the method  to derive the effective meson-baryon
Lagrangian following the work of Abu-Raddad et al~\cite{Abu-Raddad:2002pw}.
Let us start from the SU(2)$_L\times$SU(2)$_R$ NJL Lagrangian
\begin{equation}
{\cal
L}_{NJL}=\bar{q}(i\Slash{\partial}-m_0)q+\frac{G}{2}\left[(\bar{q}q)^2+(\bar{q}i\gamma_5
\vec{\tau}q)^2\right],
\label{eqn:NJLL}
\end{equation}
where $q$ is the current quark field, ${\tau}_a (a = 1,2,3)$ 
the flavor Pauli matrices, 
$G$ the NJL coupling constant with dimension of (mass)$^{-2}$ and
$m_0$ the current quark mass. In this paper, we set $m_0=0$ for
simplicity. The NJL Lagrangian is bosonized by
introducing collective meson fields as auxiliary fields in the
path-integral method. As an intermediate step, we find the following
Lagrangian:
\beq
{{\cal L}^\prime_{q\sigma\pi}} = 
\bar{q}\left( i\rlap/\partial  
-g (\sigma + i\gamma_5 \vec\tau\cdot\vec\pi) \right) q
-\frac{g^2}{2G}(\sigma^2+\vec\pi^{\; 2})\, . 
\label{Lprime}
\eeq
Here $\sigma$ and $\vec \pi$ are properly normalized 
scalar-isoscalar sigma
and pseudoscalar-isovector pion fields as generated from 
$\sigma \sim \bar q q$ and $\vec \pi \sim i\bar q \vec \tau \gamma_5q$, 
respectively, and $g$ is a meson-quark coupling constant.  

For our purpose, 
it is convenient to work in the non-linear basis~\cite{Ebert:1997hr,Ishii:2000zy,hosaka_book}.
Firstly, the meson fields are expressed as
\begin{equation}
\sigma + i\gamma_5 \vec\tau\cdot\vec\pi=f\exp\left(-\frac{i}{F_\pi}\gamma_5\vec{\tau}\cdot\vec{\Phi}\right).
\end{equation} 
where $f$ and $\Phi$ are new meson fields in the non-linear basis and
$f^2 = \sigma^2 + \vec \pi^{\; 2}$. Spontaneous breaking of chiral
symmetry is realized when $f$ takes a non-zero vacuum expectation
value $\bra f\ket=F_\pi$, which is identified with the pion decay
constant$\sim$ 93 MeV, generating the constituent quark mass
dynamically, $m_q=g F_\pi$~\cite{Ebert:1986kz}.
The non-linear Lagrangian is, then, achieved by chiral rotation 
from the current ($q$) to 
constituent ($\chi$) quark fields: 
\beq
\chi = \xi_5^\dagger q\, , \; \; \; \; 
\xi_5 
= \exp\left(\frac{i}{2F_\pi}\gamma_5\vec{\tau}\cdot\vec{\Phi}\right) \, . 
\eeq
Thus we find 
\beq
{\cal L^\prime_{\chi \sigma \pi}} = 
\bar{\chi}\left(i\rlap/\partial  
-m_q - \rlap/v - \rlap/a\gamma_5\right) \chi
-\frac{1}{2G} f^2 \, , 
\label{Lprime2}
\eeq
where 
\beq
v_\mu =  \frac{1}{2i} \left(
\xi^\dagger\del_\mu  \xi +  \xi\del_\mu  \xi^\dagger 
\right)\, , \; \; \; 
a_\mu =  \frac{1}{2i} \left(
\xi^\dagger\del_\mu  \xi -  \xi \del_\mu \xi^\dagger 
\right)\, , 
\eeq
are the vector and axial-vector currents written in terms of the 
chiral field
\beq
\xi 
= 
\exp\left(\frac{i}{2F_\pi}\vec{\tau}\cdot\vec{\Phi}\right)
.
\eeq
The Lagrangian (\ref{Lprime2}) describes not only the 
kinetic term of the quark, but also quark-meson interactions 
such as the Yukawa and the Weinberg-Tomozawa types among others.  

\begin{figure}[tbh]
\begin{center}
\includegraphics[width=5cm]{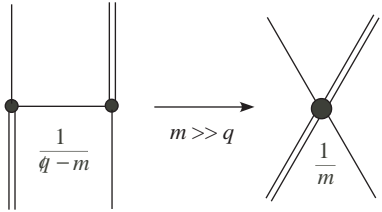}
\begin{minipage}{14cm}
   \caption{\small The quark exchange diagram(left) in the Faddeev
   approach~\cite{Buck:1992wz,Ishii:rt} and its static limit(right). }
   \label{fig:static}
 \end{minipage}
\end{center}
\end{figure}

For the description of baryons, we introduce 
diquarks and their interactions with quarks.  
We assume local interactions between a quark and a diquark
to generate the nucleon field.  
As suggested previously~\cite{Espriu:1983hu}, 
we consider two diquarks; 
one is a Lorentz scalar, isoscalar color $\bar{3}$ one
as denoted by $D$,  and the other is an 
axial-vector, isovector color $\bar{3}$ one, $D_\mu$.  
The ground state nucleon is then described as a 
superposition of the bound state of 
a quark and scalar diquark ($\equiv$ scalar channel), and 
the bound state of 
a quark and axial-vector diquark ($\equiv$ axial-vector channel).  
Hence, our microscopic Lagrangian for quarks, diquarks and 
mesons is given by~\cite{Abu-Raddad:2002pw}
\begin{eqnarray}
{\cal L} &=& \bar{\chi}(i\rlap/\del - m_q
- \rlap/v - \rlap/a\gamma_5) \chi \;-\;
\frac{g^2}{2G}f^2\;+
D^\dag (\del^2 + M_S^2)D \;+\; 
{\vec{D}^{\dag\;\mu}} 
\left[  (\del^2 + M_A^2)g_{\mu \nu} - \del_\mu \del_\nu\right]
\vec{D}^{\nu} \nonumber\\ &&
\;+\; \tilde{G} \left( \sin{\theta} \; \bar{\chi}\gamma^\mu \gamma^5
\:\vec{\tau}\cdot {\vec D}^\dag_{\mu} \;+\; \cos{\theta}
\;\bar{\chi} D^\dag\right)\; \left(\sin{\theta}\; {\vec D}_{\nu}\cdot
\vec{\tau} \: \gamma^\nu \gamma^5 \chi \;+\; \cos{\theta} \;
D \chi \right)\; .
\label{lsemibos}
\end{eqnarray}
In the last term $\tilde G$ is a coupling constant for the 
quark-diquark interaction and an angle $\theta$ 
controls the mixing ratio of the scalar and axial-vector channels
in the nucleon wave function.   
In this construction, we have assumed a local interaction between the
quark and diquarks. This stems from, for instance, the static limit of
a quark exchange between a quark and a diquark as shown in
Fig.\ref{fig:static}. In this case, due to the spin-flavor-color
structure, the interactions become attractive both for the scalar and
axial-vector diquark channels. In Eq.~(\ref{lsemibos}), a positive
$\tilde{G}$ guarantees an attractive interaction, which is the
case we consider.

Now the hadronization procedure is straightforward by the introduction
of a baryon field as an auxiliary field $B\sim
\sin\theta\vec{D}_\nu\cdot\vec{\tau}\gamma^\nu\gamma_5\chi+\cos\theta
D\chi$ and the elimination of the 
quark and diquark fields in Eq.~(\ref{lsemibos}).
The final result is written in a compact form as~\cite{Abu-Raddad:2002pw}

\begin{eqnarray}
{\cal L}_{\rm eff} &=& 
- \frac{1}{2G}f^2 
\;-\; i \;{\rm tr\; ln} (i \rlap/\del - m_q -\rlap/v - \rlap/a\gamma_5)
\;-\; \frac{1}{\tilde{G}}\; \bar{B} B \;+\; 
i\;{\rm tr \; ln} ( 1\;-\; \Box ) \, .
\label{effL}
\end{eqnarray}
Here traces are taken over space-time, color, flavor, and Lorentz indices, and 
the operator $\Box$ is defined by 
\beq
\label{eqbox}
\Box & = & 
\begin{pmatrix} {\cal A}& {{\cal F}_2}\\
 {{\cal F}_1} &{\cal S}
\end{pmatrix} \, ,
\eeq
where
\begin{subequations}
\begin{eqnarray}
{\cal A}^{\mu i,\,\nu j} & = & {\sin}^2\theta \
\bar{B} \; 
\gamma_\rho \gamma^5\;
  {\tau}_{k} \;\tilde{\Delta}^{\rho k,\,\mu i} \;
 S\; {\tau}^{j}\;
  \gamma^\nu \gamma^5\; B\;,\\ 
{\cal S}& = & {\cos}^2\theta\; \bar{B}\;\Delta
  \;S\;B\;,\;\; \;\;\;\;\\ 
({{\cal F}}_1)^{\nu j}& = & \sin{\theta} \cos{\theta}\;\bar{B}\; 
\Delta \;S\; 
 {\tau^j}\;\gamma^\nu \gamma^5\;  B\;,\\
({\cal F}_2)^{\mu i}& = & \sin{\theta} \cos{\theta} \;
\;\bar{B} \;\tilde\Delta^{\rho k,\, \mu i}\;\gamma_\rho \gamma^5\; {\tau_k} \; S\; B\;.
\end{eqnarray}
\label{eq:boxint}
\end{subequations}
The $S$, $\Delta$ and $\tilde{\Delta}$ are the quark, scalar diquark
and axial-vector diquark propagators.
The tr log can be expanded as
\begin{equation}
\tr \ln (1-\Box)=-\tr\left(\Box+\frac{\Box^2}{2}+\cdots\right).
\label{trlogbox}
\end{equation}
The first term on the right hand side describes one particle
properties of the nucleon, as it contains the nucleon bilinear form
$\bar{B}\Gamma B$, while higher order terms contain two, three and
more nucleon interaction terms.

Finally, we comment on the properties of the nucleon
field. Since we take the non-linear representation, the
transformation properties of baryons under chiral $SU(2)_L\times
SU(2)_R$  are simple. Baryons transform in the same way as quarks do
\begin{equation}
\chi\to \chi^\prime(x)=h(x)\chi(x), B(x)\to B^\prime(x)=h(x)B(x),
\end{equation}
where $h(x)$ is the non-linear function of the chiral transformations
and of the chiral field at a point $x$~\cite{hosaka_book}.
Here we note that the baryon field $B(x)$, in terms of quarks and diquarks,
is related to the nucleon wave-function in the constituent quark model
by way of 
\begin{eqnarray}
D\chi=2\phi_\rho\chi_\rho,\\
\vec{D}_\nu\cdot\vec{\tau}\gamma^\nu\gamma_5\chi=6\phi_\lambda\chi_\lambda,
\end{eqnarray} 
in the non-relativistic limit, where $\phi_\rho,\ \phi_\lambda$ and
$\chi_\rho,\ \chi_\lambda$ are the standard 3-quark spin and iso-spin 
wave-functions~\cite{hosaka_book}. If we take $\tan\theta=1/3$,
we realize the $SU(4)$ spin-flavor symmetry of the constituent quark model.
\section{The quark-diquark self-energy}
\begin{figure}[tbh]
\centering{
\includegraphics[width=5cm]{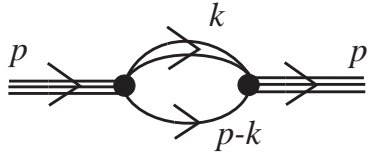}}
\begin{minipage}{14cm}
   \caption{\small A diagrammatic representation of the quark-diquark 
   self-energy. The single, double and triple lines represent the quark,
   diquark and nucleon respectively. }
   \label{fig:self}
 \end{minipage}
\end{figure}

In the first order term with respect to $\Box$ of Eq.~(\ref{trlogbox}), 
the quark-diquark self-energy corresponding to Fig.~\ref{fig:self} is given by,
\begin{eqnarray}
\calL_0= -i\tr\left[\calS+\calA^{\mu\nu}_{ij}\right]-\frac{1}{\tilde{G}}\bar{B}B.
\end{eqnarray}
Using the interaction terms in Eqs.~(\ref{eq:boxint}), we obtain
\begin{eqnarray}
\calL_0 = \cos^2\theta \bar{B}(p)\Sigma_S(p)B(p)
+\sin^2\theta\bar{B}(p)\Sigma_A(p)B(p)-\frac{1}{\tilde{G}}\bar{B}B,
\end{eqnarray}
where $\Sigma_S(p)$ and $\Sigma_A(p)$ are the nucleon self-energies 
corresponding to the scalar and axial-vector
diquark channels, respectively,
\begin{subequations}
\begin{eqnarray}
\Sigma_S(p)&=&-i N_c\int\frac{d^4k}{(2\pi)^4}\frac{1}{k^2-\ms^2}\frac{\Slash{p}-\Slash{k}+\mq}{(p-k)^2-\mq^2},\\
\Sigma_A(p)&=&-iN_c\int\frac{d^4k}{(2\pi)^4}\frac{k^\mu k^\nu/\ma^2-g^{\mu\nu}}{k^2-\ma^2}\delta_{ij}\gamma_\nu\gamma_5\tau_j
\frac{\Slash{p}-\Slash{k}+\mq}{(p-k)^2-\mq^2}\tau_i\gamma_\mu\gamma_5 . 
\end{eqnarray}
\label{SigmaSA}
\end{subequations}
These are divergent;  $\Sigma_S(p)$ is logarithmically and 
$\Sigma_A(p)$ quadratically divergent.  
In the previous
works, we employed the Pauli-Villars regularization to let 
the divergences finite. 
In the present work, however we shall 
employ the three momentum cutoff method, since 
the Pauli-Villars method is not appropriate to regularize
the quadratic divergence in $\Sigma_A$.  
The quadratic nature necessarily requires two independent 
cutoff parameters in the Pauli-Villars method, while it is sufficient to 
introduce single cutoff parameter in the three-momentum
cut-off scheme.  

Now in the rest frame of the nucleon, i.e. 
$p=(p_0, \vec{0})$, the self-energies as functions of $p_0$
can be written as, after integrating Eqs.~(\ref{SigmaSA}) over $k_0$, 
\beq
\Sigma_S(p_0)
&=&
\Sigma_S^0(p_0) + \Sigma^1_S(p_0) \gamma_0\, , \nonumber \\
\Sigma_A(p_0)
&=&
\Sigma_A^0(p_0) + \Sigma^1_A(p_0) \gamma_0\, , 
\label{defS01}
\eeq
where the coefficients are given as 
\beq
\Sigma_S^0(p_0)
&=&
- N_c\int_0^\Lambda\frac{k^2dk}{2\pi^2}
\left\{\frac{m_q}{2e_1[(p_0+e_1)^2-e_2^2]}
+\frac{m_q}{2e_2[(p_0-e_2)^2-e_1^2]}\right\}_,
\nonumber \\
\Sigma_S^1(p_0)
&=&
- N_c\int_0^\Lambda\frac{k^2dk}{2\pi^2}
\left\{\frac{-e_1}{2e_1[(p_0+e_1)^2-e_2^2]}
+ \frac{(p_0-e_2)}{2e_2[(p_0-e_2)^2-e_1^2]}\right\}_,
\label{SigmaS3}
\eeq
and 
\beq
\Sigma_A^0(p_0)
&=&
-\frac{ N_c\vec{\tau}^2}{M_A^2}\int_0^\Lambda\frac{k^2 dk}{2\pi^2}
\left\{ 
\frac{3m_q M_A^2}{2e_3[(e_3-p_0)^2-e_1^2]}\right.
\nonumber\\
& & + \; 
\left.\frac{4m_q M_A^2-m_q\left((p_0+e_1)^2-\vec{k}^2\right)}
{2e_1[(p_0+e_1)^2-e_3^2]}\right\}\, ,\nonumber\\
\Sigma_A^1(p_0)
&=&
-\frac{ N_c\vec{\tau}^2}{M_A^2}\int_0^\Lambda\frac{k^2 dk}{2\pi^2}
\left\{ 
\frac{-3e_3 M_A^2 + p_0 M_A^2 + 2 e_3^2 p_0}
{2e_3[(e_3-p_0)^2-e_1^2]}\right.
\nonumber\\
& & + \; 
\left.\frac{-2 e_1 M_A^2 - e_1 (p_0+e_1)^2 + (2p_0 + e_1)\vec k^2}
{2e_1[(p_0+e_1)^2-e_3^2]}\right\}\, .  
\label{SigmaA3}
\eeq
In the above equations $N_c$ is the number of colors, and
$
e_1 = \sqrt{\vec{k}^2+m_q^2}, 
e_2 = \sqrt{\vec{k}^2+M_S^2},
e_3 = \sqrt{\vec{k}^2+M_A^2}
$.

Now physical nucleon fields are defined such that the self-energy
becomes the nucleon propagator on the nucleon mass shell.  
This condition is implemented by expanding the self-energy around $p_0=\mn$:
\begin{eqnarray}
\bar{B} \left(\cos^2\theta\Sigma_S(p_0)+\sin^2\theta\Sigma_A(p_0)-\frac{1}{\tilde{G}}\right)B &=&
Z^{-1}\bar{B}(p_0\gamma^0-\mn)B\nonumber\\
&=&\bar{B}_{phys}(p_0\gamma^0-\mn)B_{phys},
\label{eq:rencondition}
\end{eqnarray}
where $B_{phys}=\sqrt{Z^{-1}}B$ is the properly normalized physical
nucleon field. The parameters  $Z$ and $\mn$ are the wave-function 
renormalization constant and the mass of the
nucleon. 
The parameters $Z$, $\mn$ and $\tilde{G}$ are
determined by the following conditions,
\begin{eqnarray}
\label{eq:zdeterm}
Z^{-1} &=&\left.\frac{\del\Sigma(p_0)}{\del p_0}\right|_{p_0\to M_N}=
\cos^2\theta \frac{\del\Sigma_S(\mn)}{\del
p_0}+\sin^2\theta\frac{\del\Sigma_A(\mn)}{\del p_0}_,\\
\label{eq:mdeterm}
\tilde{G}& =& (\cos^2\theta\Sigma_S(\mn)+\sin^2\theta\Sigma_A(\mn))^{-1}_.
\end{eqnarray}
Therefore, we obtain the mass of the physical nucleon by solving 
Eqs.~(\ref{eq:zdeterm}) and (\ref{eq:mdeterm}).
\section{Results}

%
To start with, we briefly discuss our parameters
which are listed in Table~\ref{tab:parameters}. We use the values in
Ref.~\cite{Abu-Raddad:2002pw} for the mass of the constituent quarks 
$m_q$, the NJL coupling constant $G$ and the cut-off mass 
$\Lambda$.  
Then $\mq$, $G$ and $\Lambda$ are determined self-consistently in 
the NJL model by solving the gap equation and  
reproducing the pion decay constant $f_\pi = 93$ 
MeV~\cite{Ebert:1986kz,Ebert:1994mf,Hatsuda:1994pi}.  
The masses of
the scalar and axial-vector diquarks, 
$M_S$ and $M_A$ may be determined 
, for instance, in the NJL model by solving the
Bethe-Salpeter equation in the corresponding diquark 
channels~\cite{Vogl:1991qt,Cahill:1987qr}. These masses have been also
calculated by QCD oriented
methods~\cite{Hess:1998sd,Maris:2002yu,Burden:1996nh}.
Results are, however, somewhat dependent on the methods. 
Here, instead of solving the BS equation rigorously, we simply choose 
a reasonable set of diquark masses. These parameters can reproduce,
for instance the mass splitting between the nucleon and delta\cite{Vogl:1991qt,Hess:1998sd}.

\begin{table}[tbh]
{\small 
\begin{center}
\begin{minipage}{14cm}
\caption{Model parameters. All parameters are in units of GeV.}\label{tab:parameters}
\end{minipage}
\begin{tabular}{ccccccc}
\vspace*{-3mm}\\
\noalign{\hrule height 0.8pt}
$\mq$ &$\ms$ &$\ma$ & $\Lambda$\\
0.39 & 0.60 & 1.05  & 0.6\\
\noalign{\hrule height 0.8pt}
\end{tabular}
\end{center}
}
\end{table}

\begin{figure}[tbh]
\begin{center}
\includegraphics[width=7cm]{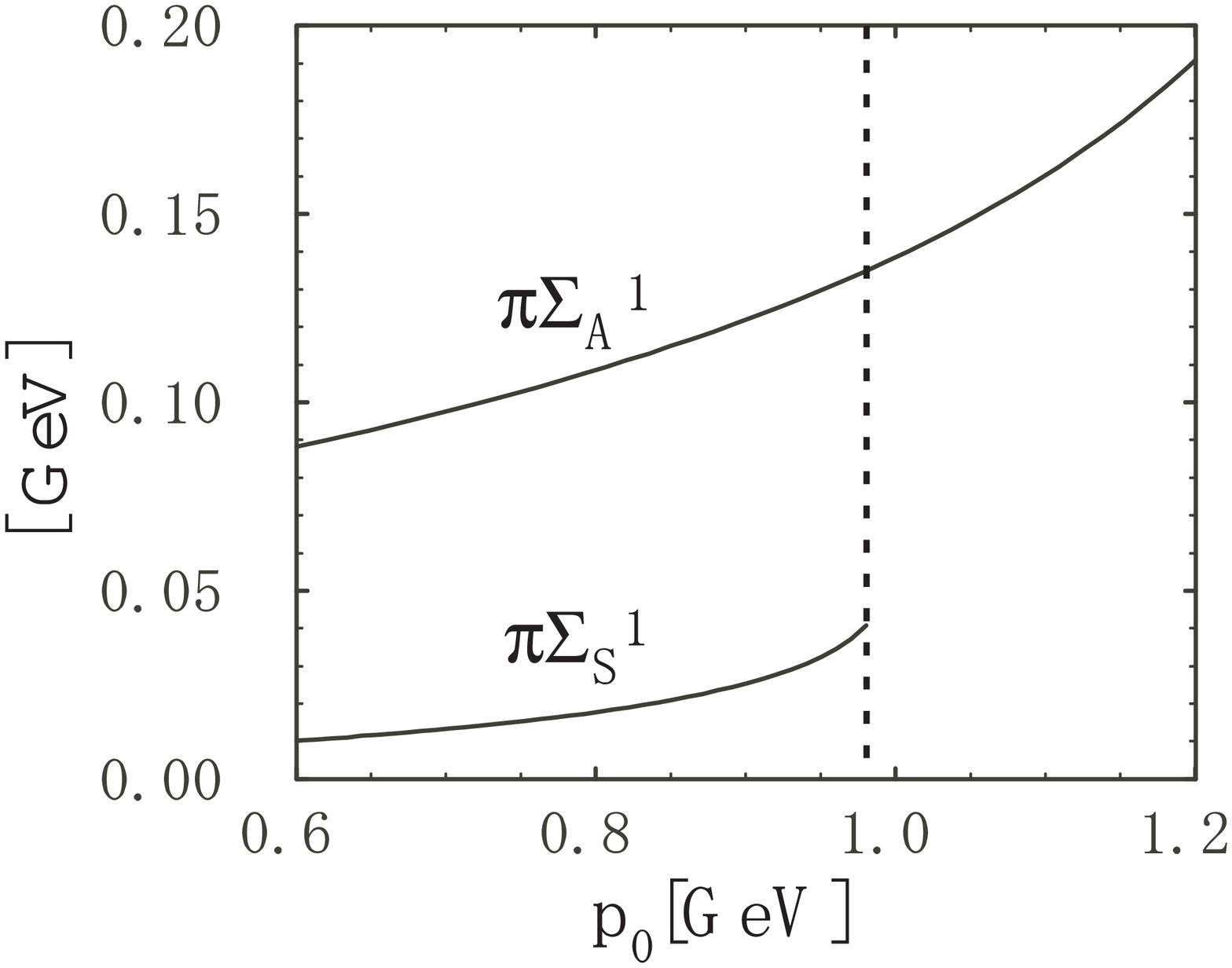}
\includegraphics[width=7cm]{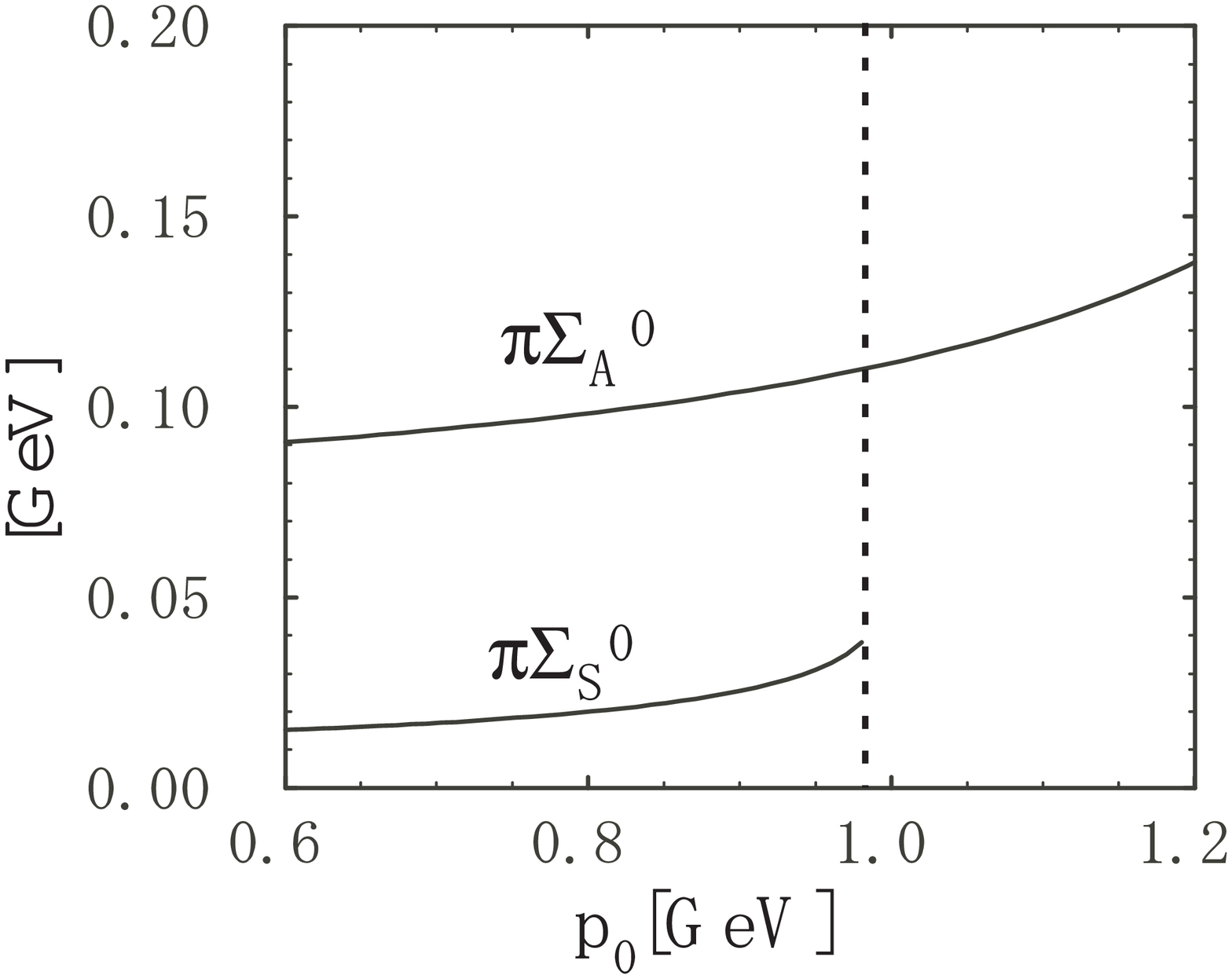}
\caption{Real parts of the self-energies $\Sigma_{S,A}^1$(left) and
$\Sigma_{S,A}^0$(right) multiplied by $\pi$. The vertical dashed lines represent the
threshold for the quark and scalar diquark channel $=m_q+M_S$.
}\label{fig:selfs}
\end{center}
\end{figure}

In Fig.~\ref{fig:selfs}, we plotted the real parts of the self-energies
Eqs.~(\ref{SigmaS3}) and (\ref{SigmaA3}) as functions of $p_0$.

We
find that both scalar and axial-vector channels are positive, meaning that
both channels contribute to the mass of the nucleon attractively, or
decrease the mass of the nucleon.
Obviously the contributions of the axial-vector diquark part is 
considerably larger than that of the scalar diquark part, reflecting
the stronger (quadratic) divergence of the former than the latter. In
Fig.~\ref{fig:selfs} we also explicitly show a threshold $m_q+M_S$. We
can not obtain the reasonable result for the larger mass of $M_N$ than
$m_q+M_S$, because this model has no confining effect. 
Recently, several works including the mimic
effects of the confinement was done\cite{Oettel:1998bk,Hellstern:1997nv,Hellstern:1997pg}. 
Although we continue this simple treatment, we should confine this
model in the region $M_N<m_q+M_S$.

\begin{figure}[tbh]
\begin{center}
\includegraphics[width=7cm]{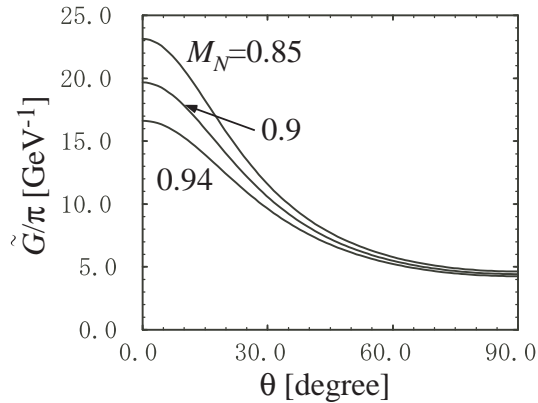}
\caption{\small The contour plot of the nucleon mass $M_N$ as a function
of the coupling constant $\tilde G$ (divided by $\pi$) and the mixing angle
$\theta$. Each lines are for $M_N$=0.94, 0.90, 0.85 [GeV] from bottom to top.
}\label{fig:gvstheta}
\end{center}
\end{figure}

In this paper, the mass of the nucleon $M_N$ is treated as a function 
of $\tilde G$ and the mixing angle $\theta$.  
In Fig.~\ref{fig:gvstheta}, 
we show the contour plot for the nucleon mass as a function of 
$\tilde{G}$ and $\theta$.
One finds that both of the scalar and axial-vector parts of the
self-energy contribute attractively to the mass of the nucleon, and
that the attraction from the axial part is larger than that from the
scalar part. 
At $\theta=0$ where only $\Sigma_S$ contributes 
to the mass of the nucleon, 
the experimental value $M_N = 0.94$ GeV is obtained when 
$\tilde G/\pi \sim 17$. In making comparisons, we note that in the present
calculation pion cloud effect is not included, which might bring
substantial contribution to nucleon properties at the quantitative
level~\cite{Thomas:1982kv,Ishii:1998tw,Hecht:2002ej}. Nevertheless, for the qualitative discussions in the
present paper, we simply compare the results with experiments directly. 

\begin{figure}[tbh]
\begin{center}
\includegraphics[width=7cm]{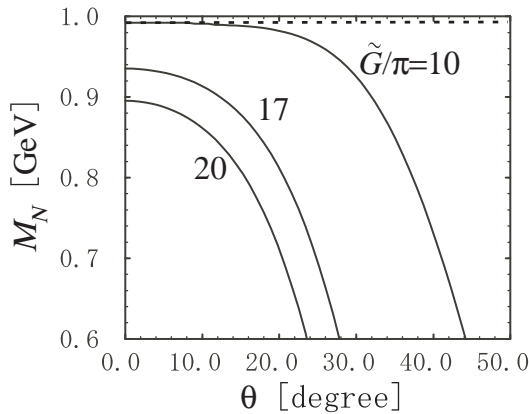}
\caption{\small The mixing angle $\theta$ dependence of the mass of
the nucleon $M_N$ for several values of the quark-diquark coupling
constant $\tilde{G}$. 
Three curves are for $\tilde{G}/\pi$=20, 17, 10
[GeV$^{-1}$] from bottom to top as indicated. 
Dashed horizontal line is the threshold $=m_q+M_S$. }\label{fig:madep}
\end{center}
\end{figure}

At $\theta = 30$ degrees, for instance, the mass of the nucleon 
is reproduced when $\tilde G/\pi \sim 9.5$.  
As the mixing angle $\theta$ increases, or the
axial-vector component in the nucleon wave-function becomes larger,
the mass of the nucleon decreases.
This behavior is also shown in Fig.~\ref{fig:madep}, where 
$\theta$ dependence is shown for fixed values of $\tilde G$.  
One sees that for larger values of $\theta$, $M_N$ 
is reproduced for smaller values of $\tilde G$, showing once again that  
the attraction from the axial-vector part is larger 
than the attraction in the scalar part. Although we do not discuss in
this paper, a finite value of $\theta$ is favored  when explaining the
isovector magnetic moments $\mu$ and isovector axial-vector coupling
constant $g_A$.

%
%

\section{Summary}
In this paper, we have studied the nucleon state in terms of a
microscopic model for hadrons, namely, 
a chiral quark-diquark model.  
The nucleon was constructed as a superposition of the 
two quark-diquark channels including the scalar and axial-vector diquarks.  
The quark-diquark model was then 
hadronized in the path-integral method to obtain 
an effective Lagrangian for the mesons and nucleon.  
The present work is an extension of the previous ones including 
only scalar diquark channel.   
Here, in order to test the validity of the method and see a role  
the axial-vector diquark channel plays, 
we investigated the mass of the nucleon.  
It was calculated through the
renormalization conditions of the nucleon self-energies.
Then we found that the mass of the nucleon is reproduced 
by choosing the mixing angle $\theta$ and the coupling constant
$\tilde G$ appropriately.  
Our result is consistent with the previous work solving the 
Faddeev equations for the three quark system in the 
NJL model~\cite{Ishii:rt}.  

The present result suggests that the quark-diquark model, 
although it is simpler than solving the three quark system 
directly, would be a practically useful model for the description 
of the nucleon.  
As advocated previously, an advantage of the present method 
is to be able to 
work out to a large extent in an analytic way preserving 
important symmetries such as gauge and chiral symmetries.  

Naturally, it is a further extension to apply the  
present method to various 
hadronic properties such as the electromagnetic 
couplings and the nuclear force.  
For some quantities such as isovector magnetic moments and 
axial-vector coupling constants, it is expected that the axial-vector channel 
 play an important role~\cite{Alkofer:2004yf,Holl:2005zi}. Furthermore, this is necessary to describe
the octet and decuplet baryons.
In addition, the axial-vector channel may play another role which
 we did not consider explicitly in the present work, for only a single 
state of the nucleon was constructed.  
If both the channels are treated as  independent degrees of
freedom, then the two nucleon states may be described as bound states
of the quark and diquarks.
This is investigated in a separate paper \cite{Nagata:2005qb}.

%

\section*{Acknowledgments}
We are grateful to Veljko Dmitrasinovic for useful suggestions and careful
reading of the manuscript.
This work was supported in part by the Sasakawa Scientific Research Grant
from The Japan Science Society.
LJA-R acknowledges the support of a joint
fellowship from the Japan Society for the Promotion of Science and the
United States National Science Foundation. 
K. N thanks D. Ebert for fruitful discussions and hospitality during
his stay in Humboldt University.

\section*{Note added}
In the numerical calculation of the self-energies in the original
version, the factor $\pi$ was missed. In order to account for the
factor correctly, the self-energies shown in Fig. 3 and the coupling
constant $\tilde{G}$ are scaled by the factor $\pi$. The results and
conclusions are not affected. This is reported in Erratum of PRC.

\end{document}